\documentclass[draft,a4paper,reqno,12pt]{amsart}
\usepackage{amssymb,euscript,bbold}
\usepackage{eepic}
\usepackage{array}
\usepackage[T1]{fontenc}
\DeclareMathOperator{\vol}{\mathrm{vol}}
\DeclareMathOperator{\spn}{\mathrm{span}}
\DeclareMathOperator{\Mat}{Mat}
\DeclareMathOperator{\slg}{SLAG}
\DeclareMathOperator{\clg}{\mathbb{C}LAG}
\newcommand{\M}{\mathsf{M}}
\newcommand{\RR}{\mathbb{R}}
\newcommand{\PP}{\mathbb{P}}
\newcommand{\U}{\mathrm{U}}
\newcommand{\SU}{\mathrm{SU}}
\renewcommand{\Sp}{\mathrm{Sp}}
\newcommand{\Spin}{\mathrm{Spin}}
\newcommand{\EE}{\mathbb{E}}
\newcommand{\MM}{\mathbb{M}}
\newcommand{\SO}{\mathrm{SO}}
\newcommand{\TT}{\mathbb{T}}
\newcommand{\Cl}{\mathrm{C}\ell}
\newcommand{\fg}{\mathfrak{g}}
\newcommand{\half}{\tfrac{1}{2}}
\newcommand{\1}{\mathbb{1}}
\newcommand{\so}{\mathfrak{so}}
\newcommand{\repre}[1]{\underline{\mathbf{#1}}}
\newcommand{\eG}{\EuScript{G}}
\newcommand{\OO}{\mathbb{O}}
\newcommand{\CC}{\mathbb{C}}
\newcommand{\HH}{\mathbb{H}}
\newcommand{\ZZ}{\mathbb{Z}}
\newtheorem{dfn}{Definition}
\newtheorem{quest}{Question}
\newtheorem{lem}{Lemma}
\newtheorem{prop}{Proposition}
\begin{document}

\title[planes, branes and automorphisms: branes in motion]{Planes,
Branes and Automorphisms\\II. Branes in motion}
\author[acharya]{BS Acharya}
\author[figueroa-o'farrill]{JM Figueroa-O'Farrill}
\author[spence]{B Spence}
\address[BSA,JMF,BS]{\begin{flushright}
Department of Physics\\
Queen Mary and Westfield College\\
Mile End Road\\
London E1 4NS, UK\end{flushright}}
\email{\{r.acharya,j.m.figueroa,b.spence\}@qmw.ac.uk}
\author[stanciu]{S Stanciu}
\address[SS]{\begin{flushright}
Theoretical Physics Group\\
Blackett Laboratory\\
Imperial College\\
Prince Consort Road\\
London SW7 2BZ, UK\end{flushright}}
\email{s.stanciu@ic.ac.uk}
\date{\today}
\begin{abstract}
We complete the classification of supersymmetric configurations of two
$\M5$-branes, started by Ohta and Townsend.  The novel configurations
not considered before are those in which the two branes are moving
relative to one another.  These configurations are obtained by
starting with two coincident branes and Lorentz-transforming one of
them while preserving some supersymmetry.  We completely classify the
supersymmetric configurations involving two $\M5$-branes, and
interpret them group-theoretically.  We also present some partial
results on supersymmetric configurations involving an arbitrary number
of $\M5$-branes.  We show that these configurations correspond to
Cayley planes in eight-dimensions which are null-rotated relative to
each other in the remaining $(2+1)$ dimensions.  The generic
configuration preserves $\tfrac{1}{32}$ of the supersymmetry, but
other fractions (up to $\tfrac{1}{4}$) are possible by restricting the
planes to certain subsets of the Cayley grassmannian.  We discuss some
examples with fractions $\tfrac{1}{32}$, $\tfrac{1}{16}$,
$\tfrac{3}{32}$, $\tfrac{1}{8}$, and $\tfrac{1}{4}$ involving an
arbitrary number of branes, as well as their associated geometries.
\end{abstract}
\maketitle

\tableofcontents

\section{Introduction}

This is Part~II in a series of papers dedicated to the
group-theoretical study of intersecting brane configurations.  In the
first paper in this series, hereafter referred to as Part~I
\cite{AFS-groups}, we outlined a complete characterisation of
configurations of multiply intersecting static branes in terms of
subgroups of $\Spin_{10}$ preserving some spinors.  In the present
paper we will consider configurations in which the branes are not
necessarily static relative to each other.  We refer the reader to
Part~I and to the references therein for a summary of the literature
on this topic as well as for the basic notation.

This paper is organised as follows.  In Section 2 we will state the
problem to be addressed, namely the determination of all
supersymmetric configurations of $m$ intersecting $\M5$-branes in
eleven-dimensional Minkowski spacetime, as well as of the fraction
$\nu$ of the supersymmetry which is preserved.  In Section 3 we solve
this problem for the special case of $m=2$ branes, and as preparation
for the following sections, we interpret the solutions
group-theoretically.  This relies heavily on the results of Part~I and
on those in \cite{OhtaTownsend}.  We find new supersymmetric
configurations corresponding to supersymmetric branes at angles which
have then been null-rotated relative to each other.  These new
configurations trace their origin to the fact that there exists a
class of singular orbits in the spinorial representation of
$\Spin_{10,1}$ whose isotropy does not leave invariant any time-like
directions.  In Section 4 we study this spinorial representation in
some detail, and collect some basic facts concerning this exotic
isotropy group.  In Section 5 we tackle the multiple intersection
problem. We start by reformulating the problem of multiple
intersections of moving branes in eight-dimensional terms.  We will
show that all such supersymmetric configurations consist of Cayley
planes in eight dimensions which have been null-rotated in the
remaining three dimensions.  These configurations generically preserve
$\frac{1}{32}$ of the supersymmetry; but by restricting the planes to
sub-orbits of the Cayley grassmannian, it is possible to find
configurations which preserve a larger fraction---the largest such
fraction being $\frac{1}{4}$.  In Section 6 we discuss some examples
of such configurations and explore their geometries.  Section 7
concludes this paper with a summary of the status of the
classification problem.

\section{The statement of the problem}

In this section we set up the problem.  Let us consider the
$\M5$-brane solution for definiteness.  Let $(x^\mu)$ denote the
eleven-di\-men\-sion\-al coordinates, where $(x^0,x^1,\ldots,x^5)$ are
coordinates along the brane and $(x^6,\ldots,x^9,x^{\natural})$ are
coordinates transverse to the brane.  Far away from the brane, the
metric is asymptotically flat, so that the Killing spinors of the
supergravity solution have constant asymptotic values $\varepsilon$,
obeying
\begin{equation}\label{eq:halfsusy}
\Gamma_{012345}\, \varepsilon = \varepsilon~,
\end{equation}
where $\varepsilon$ is a real $32$-component spinor of $\Spin_{10,1}$.
We think of $\Spin_{10,1}$ as contained in the Clifford algebra
$\Cl_{1,10}$ generated by the $\Gamma_M$.  Then the spinor
representation $\Delta$ of $\Spin_{10,1}$ can and will be identified
once and for all with one of the two inequivalent irreducible
representations of $\Cl_{1,10}$.

As in Part~I we will rewrite \eqref{eq:halfsusy} in a more convenient
notation.  Let us then fix a point $x$ in the spacetime $M$ and an
orthonormal frame $e_0,e_1,\ldots,e_9,e_\natural$ for the tangent
space at $x$.  This allows us to identify the tangent space $T_xM$
with eleven-di\-men\-sion\-al Minkowski spacetime $\MM^{10,1}$.  The
tangent space to the worldvolume of a fivebrane passing through $x$
will be a $(5,1)$-dimensional oriented (and time-oriented) subspace of
$T_xM$, or equivalently an oriented (and time-oriented) $(5,1)$-plane
in $\MM^{10,1}$.  The space of such planes is the grassmannian
$\SO^0_{10,1}/\SO^0_{5,1}\times \SO_5$, where $\SO^0$ stands for the
connected component of the identity.  If $v_0,v_1,\ldots,v_5$ is an
orthonormal basis for such a plane, we can construct a $(5,1)$-vector
$\pi = v_0 \wedge v_1 \wedge \cdots \wedge v_5$ in
$\bigwedge^{5,1}\MM^{10,1}$ which has unit norm. Conversely, to any
given unit simple $(5,1)$-vector $\pi = v_0\wedge v_1 \wedge \cdots
\wedge v_5$, we associate an oriented time-oriented $(5,1)$-plane
given by the span of the $v_i$.  The condition for supersymmetry
\eqref{eq:halfsusy} can be rewritten more generally as
\begin{equation}\label{eq:fundsusy}
\pi \cdot \varepsilon = \varepsilon~,
\end{equation}
where $\cdot$ stands for Clifford multiplication and where we have
used implicitly the isomorphism of the Clifford algebra $\Cl_{1,10}$
with the exterior algebra $\bigwedge\MM^{10,1}$.  When $\pi =
e_0\wedge e_1\wedge \cdots\wedge e_5$, equation
\eqref{eq:fundsusy} agrees with equation \eqref{eq:halfsusy}.

Because $\pi$ has unit norm, $\pi \cdot \pi = \1$, and because it has
zero trace, the subspace $\Delta(\pi)\subset\Delta$ defined by
\begin{equation*}
\Delta(\pi) \equiv \{\varepsilon\in\Delta | \pi \cdot \varepsilon =
\varepsilon\}
\end{equation*}
is sixteen-dimensional.  In other words, the $\M5$-brane preserves one
half of the supersymmetry.  Now let $\eta$ be another $(5,1)$-plane.
It also preserves one half of the supersymmetry, but both planes taken
together will in general preserve a smaller fraction $\nu$ defined to
be the number of solutions to both \eqref{eq:fundsusy} and the
analogous equation for $\eta$.  In other words, $32\nu$ is the
dimension of the subspace
\begin{equation*}
\Delta(\pi \cup \eta) \equiv \Delta(\pi) \cap \Delta(\eta) \subset
\Delta~.
\end{equation*}
Clearly we are not restricted to only two branes.  Indeed, consider
$m$ branes with tangent planes $\pi_1\equiv \pi,\pi_2,\ldots,\pi_m$.
We say that the configuration $\cup_{i=1}^m \pi_i$ is {\em
supersymmetric\/} if and only if
\begin{equation*}
\Delta(\cup_{i=1}^m \pi_i) = \bigcap_{i=1}^m \Delta(\pi_i) \neq
\{0\}~;
\end{equation*}
the fraction $\nu$ of the supersymmetry which is preserved being
determined according to
\begin{equation*}
32 \nu = \dim \Delta(\cup_{i=1}^m \pi_i)~.
\end{equation*}
A priori $\nu$ can only take the values $\tfrac{1}{32}, \tfrac{1}{16},
\tfrac{3}{32}, \ldots, \half$; although only the following fractions
are known to occur: $\tfrac{1}{32}$, $\tfrac{1}{16}$, $\tfrac{3}{32}$,
$\tfrac{1}{8}$, $\tfrac{5}{32}$, $\tfrac{3}{16}$, $\tfrac{1}{4}$ and
$\tfrac{1}{2}$.  The two fundamental questions are the following.

\begin{quest}
How can one characterise the supersymmetric configurations
$\cup_{i=1}^m \pi_i$?
\end{quest}

\begin{quest}
What fraction $\nu$ of the supersymmetry is preserved by a given
supersymmetric configuration $\cup_{i=1}^m \pi_i$?
\end{quest}

For the special case of branes which are not moving relative to each
other, so that $\pi_i = e_0 \wedge \xi_i$, where $\xi_i$ are
$5$-planes in $e_0^\perp \cong \EE^{10}$, then both questions have
been answered fully for $m=2$ in \cite{OhtaTownsend} (see also
\cite{AFS-groups}).  In \cite{AFS-cali} (see also
\cite{GibbonsPapadopoulos}) we answered the first question for all
$m$, using techniques of calibrated geometry.  In Part~I we gave a
partial answer to the second question for arbitrary $m$, recasting the
problem in terms of group theory. In the present paper we will lift
the condition that the branes be static and consider arbitrary
configurations of two or more branes.  We will answer both questions
fully for $m=2$, and will present some partial results for general
$m$.

\section{Supersymmetric intersections of two $\M5$-branes}
\label{sec:m=2}

In this section we analyse the conditions for supersymmetry of a
configuration of two intersecting $\M5$-branes.  We consider a
starting configuration in which both branes are parallel, with tangent
plane $\pi$.  This configuration is supersymmetric, preserving one
half of the supersymmetry.  Now suppose that we perform a Lorentz
transformation to one of the branes.  Then the two fundamental
questions are subsumed into one.

\begin{quest}
For which Lorentz transformations $L \in \SO^0_{10,1}$ will the
configuration with tangents $\pi$ and $L\pi$ be supersymmetric and
what fraction $\nu$ of the supersymmetry will be preserved?
\end{quest}

In order to answer this question it is first convenient to put $L$ in
a standard form which can be easily parametrised.  In the simpler case
when $L$ is a rotation in $\SO_{10}$, a normal form is given by letting
$L$ lie in a fixed maximal torus.  The conjugacy theorem for maximal
tori guarantee that this is always possible after a change of basis.
Since $\SO^0_{10,1}$ is noncompact it will not have a maximal torus,
but we can still do something similar.

\subsection{A convenient normal form for Lorentz transformations}

Suppose we are given two branes related by a Lorentz transformation
$L$ in $\SO^0_{10,1}$.  We can undo this transformation in the
following way: we first transform one of the branes so that it is no
longer moving relative to the other one, and then we simply realign
them with a rotation.  Working backwards now, we can reach the
configuration $L\pi$ from $\pi$ by first rotating $\pi \mapsto R\pi$
and then boosting.\footnote{Strictly speaking it need not be a
pure boost: we simply mean that it is {\em not\/} a rotation.}
However because of the Lorentz invariance of the worldvolume of the
branes themselves, a boost will only change the configuration if it is
in a direction normal to both branes.  In other words, if we were to
perform a boost, say, in a direction contained in $\pi \cup R\pi$,
then this boost can be undone by a further rotation in that subspace
and by a boost along one of the branes.  Hence the relative
configuration between the branes will not change.  This is not to say
that the configurations are physically indistinguishable, since when
two branes intersect, Lorentz invariance on the brane is broken and
one can detect one brane moving relative to the other brane.  However
the relative configuration of their worldvolumes does not change and
neither will its supersymmetry.  In practice, what this means is that
if $\pi = e_0 \wedge \xi$ and $R\pi = e_0 \wedge R\xi$, then for a
boost to effect any change in the configuration, it has to be along a
vector in $\xi^\perp \cap R\xi^\perp$.  Therefore for a generic
rotation $R$ in $\SO_{10}$, so that $\xi^\perp \cap R\xi^\perp =
\{0\}$, no such boost would be possible.  In other words, in order to
effect any change in the relative configuration of the branes by a
boost, it will be necessary for the initial rotation to belong to
$\SO_9$.

We are always free to choose a basis for $\MM^{10,1}$ in such a way
that the initial rotation $R\in\SO_9$ belongs to a given maximal
torus of $\SO_9$.  The canonical embedding $\SO_8\subset\SO_9$
is such that the maximal tori agree.  Therefore we can take
$R\in\SO_8$ without loss of generality.  Moreover we can always
choose our basis so that $\pi = e_0 \wedge e_1 \wedge e_3 \wedge
\cdots \wedge e_9$, and such that $R$, which is parametrised by four
angles, is given by the block-diagonal matrix
\begin{equation}\label{eq:rotation}
R(\theta) = 
\begin{pmatrix}
R_{12}(\theta_1) & & & \\
& R_{34}(\theta_2) & & \\
& & R_{56}(\theta_3) & \\
& & & R_{78}(\theta_4)
\end{pmatrix} \in \SO_8~,
\end{equation}
each $R_{jk}(\vartheta)$ being the rotation by an angle $\vartheta$ in
the 2-plane spanned by $e_j$ and $e_k$.  The angles $(\theta_i)$ are
of course only defined up to Weyl transformations.

Having rotated, we now make a Lorentz transformation normal to the
$\EE^8$ on which this $\SO_8$ acts.  In other words, we have in
effect broken the Lorentz group $\SO^0_{10,1}$ down to $\SO_8 \times
\SO^0_{2,1}$, where $\SO^0_{2,1}$ acts on the three-dimensional space
spanned by $e_0,e_9,e_\natural$.  Because $e_9$ is tangent to the
brane, boosts along $e_9$ do not alter the configuration, so that we
can restrict ourselves to those Lorentz transformations in
$\SO^0_{2,1}$ which do {\em not\/} act trivially on the $e_\natural$
direction.  The most general such element of $\SO^0_{2,1}$ is
parametrised by a vector $v = \alpha e_0 + \beta e_9$,
$\alpha,\beta\in\RR$,
\begin{equation*}
S(v) = \exp \left( \alpha \Sigma_{0\natural} + \beta
\Sigma_{9\natural}\right)~,
\end{equation*}
where $\Sigma_{\mu\nu}$ are the generators of $\so_{2,1}$.

We can distinguish three different types of transformations $S(v)$
depending on whether $v$ is time-like, space-like or null.  If $v$ is
space-like we are basically doing a rotation in the plane $v \wedge
e_\natural$.  Since $v$ belongs to both $\pi$ and $R\pi$, we can
change basis from $\{e_0,e_9,e_\natural\}$ to $\{e_0, v/|v|,
e_\natural\}$ and the transformation $L = R(\theta) S(v)$ is simply a
rotation in $\SO_{10}$, which was treated in detail in
\cite{OhtaTownsend,AFS-groups}.  If $v$ is time-like, then we can
change basis from $\{e_0,e_9,e_\natural\}$ to
$\{v/|v|,e_9,e_\natural\}$.  $S(v)$ now corresponds to a pure boost in
the $e_\natural$ direction.  We will see below that the transformation
$L = R(\theta) S(v)$ will only preserve some supersymmetry if the
boost parameter is zero, so that $S(v) = \1$.  Therefore this case
once again falls into the analysis in \cite{OhtaTownsend,AFS-groups}.
Finally, we consider the case of $v$ null.  In this case, $S(v)$
leaves $v$ invariant, and hence corresponds to a {\em null rotation\/}
(see, e.g., \cite{PenroseRindler}).  As will see, this case will give
rise to new supersymmetric configurations.

To summarise, we can always choose basis such that the plane $\pi =
e_0\wedge e_1 \wedge e_3 \wedge \cdots \wedge e_9$ and such that the
Lorentz transformation $L$ takes the form
\begin{equation}\label{eq:normalform}
L = \left( \begin{array}{c|c}
    R & \\ \hline
      & S
    \end{array}\right) \in \SO_8 \times \SO^0_{2,1} \subset
    \SO^0_{10,1}~,
\end{equation}
where $R$ is in the maximal torus defined in \eqref{eq:rotation} and
where $S$ is either a rotation in the $e_9\wedge e_\natural$ plane, a
boost in the $e_\natural$ direction, or a null rotation.  We now
proceed to analyse the supersymmetry of a configuration of two branes
with tangent planes $\pi$ and $L\pi$, with $\pi$ and $L$ given above.

\subsection{Conditions for supersymmetry}

We are interested in solving equation \eqref{eq:fundsusy}
simultaneously for the tangent planes $\pi$ and $L\pi$.  Let
$\widehat{L}$ denote any one of the two possible lifts to
$\Spin_{10,1}$ of the Lorentz transformation $L\in\SO^0_{10,1}$.  Then
equation \eqref{eq:fundsusy} for $L\pi$ can be written as follows:
\begin{equation}\label{eq:susylpi}
\widehat{L} \cdot \pi \cdot \widehat{L}^{-1} \cdot \varepsilon =
\varepsilon~.
\end{equation}
Up to a sign, $\widehat{L}$ is given by
\begin{equation*}
\widehat{L} = \exp \left( \half \theta_1 \Gamma_{12} +
\half \theta_2 \Gamma_{34} + \half \theta_3 \Gamma_{56} + 
\half \theta_4 \Gamma_{78} + \half \alpha \Gamma_{0\natural} + \half
\beta \Gamma_{9\natural}\right)~,
\end{equation*}
from which it follows that since $\pi = e_0 \wedge e_1 \wedge e_3
\wedge \cdots \wedge e_9$, 
\begin{equation}\label{eq:onlytwo}
\pi \cdot \widehat{L}^{-1} = \widehat{L} \cdot \pi~.
\end{equation}
Plugging this into \eqref{eq:susylpi}, and using \eqref{eq:fundsusy},
we find that \eqref{eq:susylpi} follows from \eqref{eq:fundsusy} and
\begin{equation}\label{eq:OT2}
\widehat{L}^2 \cdot \varepsilon = \varepsilon~,
\end{equation}
with the same equation resulting for the other possible lift
$-\widehat{L}$.

In order to analyse this equation, it is convenient to break up
$\widehat{L}$ as in \eqref{eq:normalform}, $\widehat{L} = \widehat{R}
\widehat{S}$.  Then equation \eqref{eq:OT2} becomes
\begin{align*}
\varepsilon &= \widehat{R}^2 \cdot \widehat{S}^2 \cdot \varepsilon\\
&= \exp \left( \theta_1 \Gamma_{12} + \theta_2 \Gamma_{34} + \theta_3
\Gamma_{56} + \theta_4 \Gamma_{78} \right) \cdot \exp \left( \alpha
\Gamma_{0\natural} + \beta \Gamma_{9\natural} \right) \cdot
\varepsilon~.
\end{align*}
The first exponential expands to
\begin{equation*}
\widehat{R}^2 = \left(\cos\theta_1 + \sin\theta_1 \Gamma_{12}\right)
\cdot \cdots \cdot \left(\cos\theta_4 + \sin\theta_4
\Gamma_{78}\right)~,
\end{equation*}
which is an element in the maximal torus of $\Spin_8$.  For the
second exponential we have to distinguish three cases:
\begin{enumerate}
\item \underline{$\alpha^2 < \beta^2$}.  In this case, let $\gamma = +
      \sqrt{\beta^2 - \alpha^2}$. The exponential becomes
\begin{equation*}
\widehat{S}^2 = \cos\gamma + \sin\gamma  \left(
\frac{\alpha}{\gamma} \Gamma_{0\natural} + \frac{\beta}{\gamma}
\Gamma_{9\natural}\right)~.
\end{equation*}
Notice that in this case, the matrix in parenthesis obeys
\begin{equation*}
\left(\frac{\alpha}{\gamma} \Gamma_{0\natural} + \frac{\beta}{\gamma}
\Gamma_{9\natural}\right)^2 = -\1~,
\end{equation*}
whence it is a complex structure just like $\Gamma_{12}$,
$\Gamma_{34}$, $\Gamma_{56}$ and $\Gamma_{78}$ with which it
commutes.  This means that $\widehat{L}^2 = \widehat{R}^2
\widehat{S}^2$ belongs to the maximal torus of $\Spin_{10}$.  This case
was studied in \cite{OhtaTownsend} and \cite{AFS-groups} and we will
have nothing more to add here.

\item \underline{$\alpha^2 > \beta^2$}.  In this case, let $\gamma = +
      \sqrt{\alpha^2 - \beta^2}$. The exponential becomes
\begin{equation*}
\widehat{S}^2 = \cosh\gamma + \sinh\gamma \left(
\frac{\alpha}{\gamma} \Gamma_{0\natural} + \frac{\beta}{\gamma}
\Gamma_{9\natural}\right)~.
\end{equation*}
Defining $\delta = - \sqrt{-1}\gamma$ and noticing that $\sinh\gamma =
\sqrt{-1} \sin\delta$, we can rewrite $\widehat{S}^2$ as an imaginary
rotation
\begin{equation*}
\widehat{S}^2 = \cos\delta + \sin\delta \left(
\frac{\alpha}{\delta} \Gamma_{0\natural} + \frac{\beta}{\delta}
\Gamma_{9\natural}\right)~,
\end{equation*}
where the matrix in parenthesis is also a complex structure
\begin{equation*}
\left(\frac{\alpha}{\delta} \Gamma_{0\natural} + \frac{\beta}{\delta}
\Gamma_{9\natural}\right)^2 = -\1~.
\end{equation*}
This means that formally $\widehat{L}^2 = \widehat{R}^2 \widehat{S}^2$
belongs to the maximal torus of $\Spin_{10}$, but where one of the
angles is pure imaginary.  This will allow us to use the results of
\cite{OhtaTownsend} to treat this case.  It was shown in
\cite{OhtaTownsend} (see also \cite{AFS-cali,AFS-groups}) that this
configuration is supersymmetric if and only if the sum of the angles
is zero modulo $2\pi$.  In \cite{OhtaTownsend} this follows from an
explicit calculation which only uses the fact that the four complex
structures in $\widehat{R}^2$ and the one in $\widehat{S}^2$ all
commute.  Since the angles can now be complex, their sum vanishes
(modulo $2\pi$) provided that both the real and imaginary parts of
their sum vanish.  Therefore $\delta$, being the only imaginary angle,
has to vanish, and $\widehat{L}^2 = \widehat{R}^2$ which is a rotation
in the maximal torus of $\Spin_8$. This therefore reduces to the cases
studied before, and will not studied further in this paper.

\item \underline{$\alpha^2 = \beta^2$}.  In this case, the exponential
      truncates to a linear term:
\begin{equation*}
\widehat{S}^2 = \1 + \alpha \Gamma_{0\natural} + \beta
\Gamma_{9\natural}~.
\end{equation*}
This case will yield the constructions of new supersymmetric
configurations of intersecting branes in motion, and we now turn to
its detailed analysis.
\end{enumerate}

\subsection{Supersymmetry and null rotations}

Last, but not least, we consider Case 3.  Now we have
\begin{equation*}
\widehat{L}^2 = \widehat{R}^2 \cdot \left( \1 + \widehat{N} \right)~,
\end{equation*}
where $\widehat{N}$ denotes the nilpotent matrix
\begin{equation*}
\widehat{N} \equiv \alpha \Gamma_{0\natural} + \beta
\Gamma_{9\natural}~,
\end{equation*}
obeying $\widehat{N}^2 = 0$.  Equation \eqref{eq:OT2} becomes
\begin{equation*}
\left( \widehat{R}^2 - \1 + \widehat{N} \right) \cdot \varepsilon =
0~.
\end{equation*}
This equation means that $\varepsilon$ lies in the kernel of a linear
transformation with a Jordan--Chevalley decomposition consisting of a
nilpotent piece $\widehat{N}$ and a semisimple piece ($\widehat{R}^2 -
\1$), which commute with each other.  It follows that $\varepsilon$
must be annihilated by both pieces simultaneously.  Therefore for two
branes related by a rotation $R$ together with a null rotation $(\1 +
N)$ in a perpendicular plane, the conditions for supersymmetry become
the following three equations:
\begin{equation}\label{eq:three}
\pi \cdot \varepsilon = \varepsilon~,\qquad
\widehat{R}^2 \cdot \varepsilon = \varepsilon \qquad \text{and} \qquad
\widehat{N} \cdot \varepsilon = 0~.
\end{equation}
The first two equations are the conditions for two rotated branes to
be supersymmetric.  They were originally studied in
\cite{OhtaTownsend} who classified all possible configurations.  In
their nomenclature they correspond to those rotations with four angles
or less.  In \cite{AFS-groups} we showed that rotations
$\widehat{R}^2$ for which these equations have some solutions must
belong to the maximal torus of $\Spin_7$ (equivalently the maximal
torus of its subgroup $\SU_4$) in $\Spin_8$.  Such configurations
generically preserve a fraction $\nu=\frac{1}{16}$ of the
supersymmetry, but one can preserve a larger fraction by specialising
to a descending chain of subgroups.  The total fraction $\nu$, once we
have imposed the third equation in \eqref{eq:three}, will be further
halved, because only half of those spinors which satisfy the first two
equations also satisfy the third.  Let us see this in more detail.

To understand the third equation, let us introduce the ``contracting
homotopy'' $K = \alpha \Gamma_{0\natural} - \beta \Gamma_{9\natural}$,
which satisfies
\begin{equation}\label{eq:homotopy}
K \cdot \widehat{N} + \widehat{N} \cdot K = 4 \alpha^2 \1~.
\end{equation}
Applying both sides of the equation to $\varepsilon$ we see that
\begin{equation*}
\widehat{N} \cdot \varepsilon = 0 \implies
\varepsilon = \widehat{N} \cdot \left( \widehat{K} \cdot
\varepsilon \right)~,
\end{equation*}
where we have introduced $\widehat{K} \equiv \frac{1}{4\alpha^2} K$.
A similar result holds for spinors which are annihilated by
$\widehat{K}$.  As a consequence of equation \eqref{eq:homotopy}, the
spinor representation $\Delta$ has a vector space decomposition:
\begin{equation}\label{eq:Hodgedec}
\Delta = \ker\widehat{N} \oplus \ker\widehat{K}~,
\end{equation}
and $\widehat{K}$ provides an isomorphism $\ker\widehat{N} \to \ker
\widehat{K}$.  In other words, $\dim\ker\widehat{N} = \dim\ker
\widehat{K} = 16$.  Because $\widehat{R}^2$ commutes with
$\widehat{N}$ and with $\widehat{K}$, it respects and hence refines
the decomposition.

We now describe the action of $\widehat{R}^2$ on $\Delta$.  Because
$\widehat{R}^2$ belongs to (the maximal torus of) $\Spin_8$, we first
realise $\Delta$ as a representation of $\Spin_8 \subset\Cl_8$.  As
will be shown later, in terms of representation of $\Cl_8$, $\Delta =
\Delta' \otimes \RR^2$, where $\Delta'$ is the irreducible real
representation of $\Cl_8$ and $\RR^2$ is two copies of the trivial
representation.  In terms of $\Spin_8$, $\Delta'$ breaks up further as
$\Delta'_+ \oplus \Delta'_-$, which are the representations of
definite chirality; although we will not need this further
decomposition in the sequel.

Under the action of $\widehat{R}^2$, $\Delta'$ breaks up as follows:
\begin{equation*}
\Delta' = \bigoplus_\vartheta n_\vartheta \Delta'_\vartheta~,
\end{equation*}
where $\vartheta$ are angles and the multiplicity $n_\vartheta$ is a
positive integer.  For $\vartheta\neq 0,\pi$,
$\widehat{R}^2$ restricts to $\Delta'_\vartheta$ as a $2\times2$
matrix of the form
\begin{equation*}
\begin{pmatrix}
\phantom{-}\cos\vartheta & \sin\vartheta\\
-\sin\vartheta & \cos\vartheta
\end{pmatrix}~.
\end{equation*}
For $\vartheta=0,\pi$, $\Delta'_0$ and $\Delta'_\pi$ are
one-dimensional with $\widehat{R}^2$ restricting to $\pm 1$
respectively.  A closer look at the weights of the half-spin
representations of $\Spin_8$ shows that if a weight appears, then so
does its negative; whence $n_0$ and $n_\pi$ are actually even.
Applying this to $\Delta$, we have the following decomposition
\begin{equation*}
\Delta = \bigoplus_\vartheta 2n_\vartheta \Delta'_\vartheta~.
\end{equation*}
Let $\Delta_0$ denote the $2n_0$-dimensional subspace of $\Delta$
defined by those $\varepsilon\in\Delta$ obeying the second equation in
\eqref{eq:three}.  As we have just seen, the dimension of $\Delta_0$
is divisible by $4$.

Because $\widehat{N}$ and $\widehat{K}$ commute with $\widehat{R}^2$,
they preserve $\Delta_0$.  Let $\widehat{N}_0$ and $\widehat{K}_0$
denote their restrictions to $\Delta_0$.  Equation \eqref{eq:homotopy}
again implies that we can decompose
\begin{equation*}
\Delta_0 = \ker\widehat{N}_0 \oplus \ker \widehat{K}_0~,
\end{equation*}
with $\widehat{K}_0$ giving an isomorphism $\ker\widehat{N}_0 \to
\ker\widehat{K}_0$; whence they have the same dimension: $\half$ the
dimension of $\Delta_0$.

Similarly, from \eqref{eq:onlytwo}, $\pi$ preserves $\Delta_0$ and
also this decomposition.  Let $\pi_0$ denote the restriction of $\pi$
to $\Delta_0$.  Since $\pi_0 \cdot \pi_0 = \1$, $\pi_0$ decomposes
$\Delta_0$ as
\begin{equation*}
\Delta_0 = \Delta_0^+ \oplus \Delta_0^-~,
\end{equation*}
and also the subspace $\ker\widehat{N}_0$ as
\begin{equation*}
\ker\widehat{N}_0 = (\ker\widehat{N}_0)^+ \oplus
(\ker\widehat{N}_0)^-~,
\end{equation*}
in the obvious way.  Equation \eqref{eq:three} says that $\varepsilon$
belongs to $(\ker\widehat{N}_0)^+$, and we would like to compute its
dimension.  We will show that it is $\tfrac{1}{4}$ the dimension of
$\Delta_0$.

To see this consider $\Gamma_{12}$.  It commutes with $\widehat{R}^2$,
$\widehat{N}$ and $\widehat{K}$ and anticommutes with $\pi$.
Therefore it maps $\Delta_0^+$ isomorphically to $\Delta_0^-$ and in
particular $(\ker\widehat{N}_0)^+$ isomorphically to
$(\ker\widehat{N}_0)^-$.  It follows that these latter subspaces
have the same dimension: $\half$ the dimension of $\ker\widehat{N}_0$.
In summary,
\begin{equation*}
\dim(\ker\widehat{N}_0)^+ = \half \dim\ker\widehat{N}_0 = \tfrac14
\dim\Delta_0~.
\end{equation*}

The second equation in \eqref{eq:three} says that $\varepsilon$
belongs to $\Delta_0$, whereas the third equation forces it to belong
to $\ker\widehat{N}_0$.  The first equation in \eqref{eq:three}
further constraints $\varepsilon$ to live in $(\ker\widehat{N}_0)^+$.
In other words, the number of spinors satisfying all three equations
in \eqref{eq:three} is one half the number satisfying the first two.
From the known results \cite{OhtaTownsend} about the solutions to the
first two equations, we can immediately write down the possible
fractions in terms of the codimension of the spatial intersection of
the two branes (see \cite{AFS-groups}):

\begin{table}[h!]
\centering
\setlength{\extrarowheight}{5pt}
\begin{tabular}{|c|>{$}l<{$}|}
\hline
Codimension $d$& \multicolumn{1}{c|}{Fractions $\nu$}\\
\hline\hline
4 & \tfrac{1}{32} \to \tfrac{1}{16} \to \tfrac{3}{32}
\to \tfrac{1}{8}\\
3 & \tfrac{1}{16}\\
2 & \tfrac{1}{8}\\
0 & \tfrac{1}{4}\\[5pt] \hline
\end{tabular}
\vspace{8pt}
\caption{Fractions of supersymmetry appearing in configurations of two
null-rotated $\M5$-branes, in terms of the codimension of the
intersection.  Arrows indicate progressive specialisation.}
\label{tab:OTNR}
\end{table}

\subsection{Group-theoretical analysis}

We now interpret the above results in terms of group theory, as was
done in Part~I for the solutions found in \cite{OhtaTownsend}.  We
will be brief because, as we have just seen, the new solutions are
obtained by null-rotating some of the supersymmetric configurations
consisting of two intersecting $\M5$-branes at angles, and the group
theory for those configurations has been discussed in detail in
Part~I.

Not every intersecting brane configuration can be null rotated to
obtain a different supersymmetric configuration.  In the notation of
\cite{OhtaTownsend} we need to restrict ourselves to configurations
with at most four angles, whereas in the notation of Part~I, the
codimension must be at most four.

\begin{figure}[h!]
\centering
\setlength{\unitlength}{0.01in}
\begin{picture}(450,280)(-1,-10)
\put(0,240){\makebox(0,0)[lb]{$\Spin_7$}}
\path(80,250)(45,250)
\path(53,252)(45,250)(53,248)
\put(60,250){\makebox(0,0)[lb]{$\bullet$}}
\put(85,240){\makebox(0,0)[lb]{$\SU_4$}}
\path(155,250)(120,250)
\path(128,252)(120,250)(128,248)
\put(160,240){\makebox(0,0)[lb]{$\Sp_2$}}
\path(225,250)(190,250)
\path(198,252)(190,250)(198,248)
\put(205,250){\makebox(0,0)[lb]{$\bullet$}}
\put(230,240){\makebox(0,0)[lb]{$\Sp_1\times\Sp_1$}}
\path(340,250)(305,250)
\path(313,252)(305,250)(313,248)
\put(345,240){\makebox(0,0)[lb]{$\Sp_1$}}
\path(375,250)(410,250)
\path(383,252)(375,250)(383,248)
\put(390,250){\makebox(0,0)[lb]{$\bullet$}}
\put(415,240){\makebox(0,0)[lb]{$\U_1$}}
\path(20,200)(20,236)
\path(22,228)(20,236)(18,228)
\path(100,140)(100,236)
\path(102,228)(100,236)(98,228)
\put(14,180){\makebox(0,0)[lb]{$G_2$}}
\path(95,140)(35,180)
\path(42.766,177.226)(35,180)(40.547,173.898)
\put(65,160){\makebox(0,0)[lb]{$\bullet$}}
\put(85,120){\makebox(0,0)[lb]{$\SU_3$}}
\path(100,80)(100,116)
\path(102,108)(100,116)(98,108)
\path(110,80)(265,236)
\path(257.943, 231.735)(265, 236)(260.78, 228.915)
\put(85,60){\makebox(0,0)[lb]{$\SU_2$}}
\path(100,25)(100,55)
\path(102,47)(100,55)(98,47)
\put(88,3){\makebox(0,0)[lb]{$\{1\}$}}
\end{picture}
\caption{Subgroups of $\Spin_8$ associated with intersecting
brane configurations which can be null-rotated while preserving some
supersymmetry. Arrows represent embeddings, and those adorned with a
$\bullet$ are such that the maximal tori agree.}\label{fig:groups}
\end{figure}
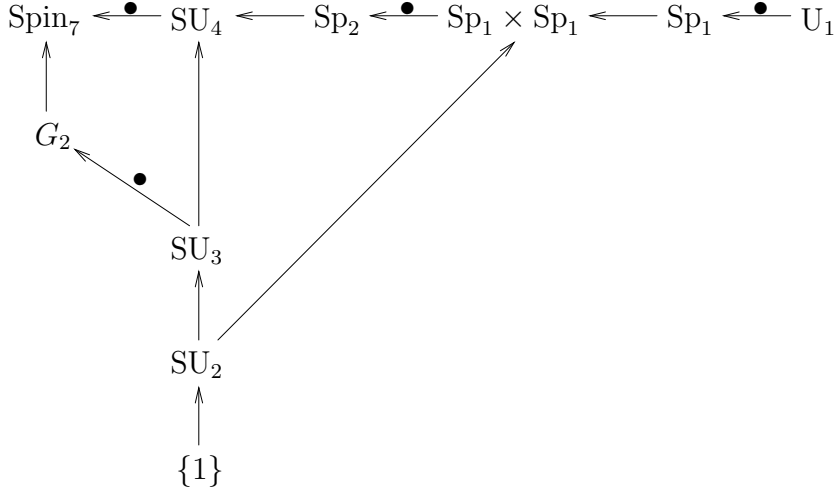

These configurations were shown in Part~I to correspond to subgroups
$G\subset\Spin_8$ which preserve some spinor.  More precisely,
$\widehat{R}^2 \in \TT(G)$, the maximal torus of $G$.  A list of
possible subgroups is displayed in Figure \ref{fig:groups}.  The first
row consists of subgroups of $\Spin_8$, the second of subgroups of
$\Spin_7$, the third of $\Spin_6$ and the fourth of $\Spin_4$.  The
first row is nothing but the spin series: $\Spin_7 \supset \Spin_6
\cong \SU_4 \supset \Spin_5 \cong \Sp_2 \supset \Spin_4 \cong \Sp_1
\times \Sp_1 \supset \Spin_3 \cong \Sp_1 \supset \Spin_2 \cong \U_1$.

The two $\M5$-branes are related by $\TT(G)$ in the above figure as
well as by a null rotation, which forms a subgroup
$\RR\subset\Spin_{10,1}$. Therefore the branes are
($\TT(G)\times\RR$)-related.  As discussed for example in Part~I, the
groups $G$ relating the branes are subgroups of $\SU_5$ or $\Spin_7$,
which, together with their intersection $\SU_4$, are the subgroups of
$\Spin_{10}$ which leave a spinor invariant.  In the more general
setup of this paper, we expect the groups $\TT(G) \times \RR$ to be
subgroups of the isotropy groups of spinors in $\Spin_{10,1}$.  In
fact, as we shall see in the next section this is indeed the case.
The group in question is a non-reductive $30$-dimensional Lie subgroup
of $\Spin_{10,1}$ isomorphic to $\Spin_7 \ltimes \RR^9$, where
$\Spin_7 \subset \Spin_8$ leaves the vector representation
irreducible, and where $\RR^9$ acts by null rotations.

\section{Null rotations and an exotic spinor isotropy group}

As preparation for our discussion of the multiple intersection problem
in the next section, we collect some basic facts about
eleven-dimensional spinors and in particular their `exotic' isotropy.
It will be convenient for some of the calculations to work in a
specific realisation for the Clifford algebra $\Cl_{1,10}$, so we
start by discussing this.  We then turn our attention to the isotropy
subgroup $(\Spin_7 \ltimes \RR^9) \subset \Spin_{10,1}$.  The $\RR^9$
subgroup acts as null rotations and we collect some results about
them.

\subsection{An explicit real realisation of $\Cl_{1,10}$}

Because as a real associative algebra, $\Cl_{1,10} \cong
\Mat_{32}(\RR)\oplus\Mat_{32}(\RR)$, there are two inequivalent
irreducible representations: both real and of dimension 32.  Choosing
a set of generators $\Gamma_0, \Gamma_1, \ldots, \Gamma_9,
\Gamma_\natural$ for $\Cl_{1,10}$, their product
$\Gamma_{012\cdots9\natural}$ commutes with all $\Gamma_M$ and squares
to one.  Hence by Schur's lemma it is $\pm\1$ on an irreducible
representation.  We choose $\Delta$ to be the one on which
$\Gamma_{012\cdots9\natural}$ takes the value $-\1$.  This means that
$\Gamma_0 = \Gamma_{1\cdots9\natural}$, where $\Gamma_1, \ldots,
\Gamma_9, \Gamma_\natural$ generate $\Cl_{0,10}$.  The Clifford algebra
$\Cl_{0,10}$ is isomorphic to $\Cl_8\otimes\Cl_{0,2}$, where the
isomorphism is given explicitly as follows in terms of generators.
Let $\Gamma'_i$ for $i=1,2,\ldots,8$ denote the generators for
$\Cl_8$ and let $\Gamma''_1$ and $\Gamma''_2$ denote the generators
for $\Cl_{0,2}$.

The $\Gamma'_i$ can be constructed explicitly in terms of octonions.
The construction of the two irreducible representations of $\Cl_7$ in
terms of octonions is well known: see, for example, \cite{LM}.
Let $\{o_i\}$, $i=1,\ldots,7$, be a set of imaginary octonion units.
Then left $L_i$ and right multiplication $R_i$ by $o_i$ on $\OO$
define the two inequivalent irreducible representations of the
Clifford algebra $\Cl_7$.  Either representation can be used in order to
build the unique irreducible representation of $\Cl_8$, we choose
$L_i$.  We define
\begin{equation*}
\Gamma'_i = \begin{pmatrix}
	    0 & L_i \\ L_i & 0
	    \end{pmatrix}
\quad\text{for $i=1,\ldots,7$; and}\quad
\Gamma'_8 = \begin{pmatrix}
	    0 & -\1 \\ \1 & 0
	    \end{pmatrix}~.
\end{equation*}
This gives rise to a manifestly real $16$-dimensional representation.

As associative algebras, $\Cl_{0,2}\cong\Mat_2(\RR)$, so we can choose
a basis
\begin{xalignat*}{2}
\Gamma''_1 &= \sigma_1 = \begin{pmatrix}0& 1\\ 1 &
0\end{pmatrix}\qquad & \Gamma''_2 &= \sigma_3 = \begin{pmatrix}1& 0\\
0 & -1\end{pmatrix}\\
\therefore\qquad \Gamma''_3 & = \Gamma''_1\Gamma''_2 = -i\sigma_2 =
\begin{pmatrix}0& -1\\ 1 & 0\end{pmatrix}~.
\end{xalignat*}
Then the generators of $\Cl_{0,10}$ are given by
\begin{align*}
\Gamma_i &= \Gamma'_i \otimes \Gamma''_3\qquad\text{for
$i=1,2,\ldots,8$}\\
\Gamma_9 &= \1 \otimes \Gamma''_1\\
\Gamma_\natural &= \1 \otimes \Gamma''_2\\
\therefore\qquad \Gamma_0 &= \Gamma'_9 \otimes \Gamma''_3~,
\end{align*}
where $\Gamma'_9 \equiv \Gamma'_1\Gamma'_2\cdots\Gamma'_8$.  This
decomposition induces an isomorphism $\Delta \cong \RR^{16} \otimes
\RR^2$, so that we can write our spinors as two-component objects,
each component being a sixteen-dimensional real spinor of
$\Cl_8$. Therefore in terms of $\Cl_8$ generators we have
\begin{alignat}{2}\label{eq:gammas}
\Gamma_0 &= \begin{pmatrix}0 & -\Gamma'_9 \\ \Gamma'_9 &
0\end{pmatrix} \qquad & \Gamma_i &=
\begin{pmatrix}0 & -\Gamma'_i\\ \Gamma'_i &
0\end{pmatrix}\quad\text{for $i=1,2,\ldots,8$}\notag\\
\Gamma_9 &= \begin{pmatrix}0 & \1 \\ \1 & 0\end{pmatrix} \qquad
& \Gamma_\natural &= \begin{pmatrix}\1 & 0\\ 0 & -\1\end{pmatrix}~.
\end{alignat}

The standard basis for the Lie algebra $\so_{10,1} \subset \Cl_{1,10}$
is given by $\Sigma_{MN} = - \half \Gamma_{MN}$, which in the chosen
realisation becomes
\begin{alignat*}{2}
\Sigma_{ij} &= \begin{pmatrix}\Sigma'_{ij} & 0 \\ 0 &
\Sigma'_{ij}\end{pmatrix}\qquad & \Sigma_{i9} &=
\begin{pmatrix}\half \Gamma'_i & 0 \\ 0 & -\half
\Gamma'_i\end{pmatrix}\\
\Sigma_{i\natural} &= \begin{pmatrix}0 & -\half\Gamma'_i\\ -\half
\Gamma'_i & 0\end{pmatrix}\qquad & \Sigma_{0i} &=
\begin{pmatrix}\half \Gamma'_9\Gamma'_i & 0 \\ 0 & \half
\Gamma'_9\Gamma'_i\end{pmatrix}\\
\Sigma_{09} &= \begin{pmatrix}\half\Gamma'_9 & 0\\ 0 & -\half
\Gamma'_9\end{pmatrix}\qquad & \Sigma_{0\natural} &=
\begin{pmatrix}0 & -\half \Gamma'_9 \\ -\half \Gamma'_9 & 0
\end{pmatrix}\\
\Sigma_{9\natural} & = \begin{pmatrix}0 & \half \1 \\ -\half \1 & 0
\end{pmatrix}~,&&
\end{alignat*}
where $\Sigma'_{ij}$ are the generators of $\so_8\subset\Cl_8$.
In particular, notice that as mentioned above, the representation
$\Delta$ breaks up under $\Spin_8$ as two copies each of the half-spin
representations.  In a more traditional language, under the embedding
$\Spin_{10,1} \supset \Spin_8$,
\begin{equation}\label{eq:spin8branching}
\repre{32} = 2\,\repre{8}_{\mathbf{s}}\oplus
2\,\repre{8}_{\mathbf{c}}~.
\end{equation}

\subsection{Spinor isotropies}

As was done in Part~I, in order to construct supersymmetric
configurations involving more than two branes---indeed an arbitrary
number---we can consider branes which are related by subgroups $G$ of
the isotropy group of a spinor.  This prompts the following question:

\begin{quest}
Which are the possible subgroups of $\Spin_{10,1}$ which leave
invariant a spinor?
\end{quest}

This question is intimately linked to the orbit decomposition under
$\Spin_{10,1}$ of its spinor representation $\Delta$.  One way to
study this problem is to define a $\Spin_{10,1}$ invariant function on
$\Delta$ which distinguishes the orbits.  Bryant \cite{Bryant-spinors}
defines a quartic polynomial invariant on $\Delta$.  This quartic
polynomial is nothing but the norm of the vector associated to 
the spinor.  In other words, if $\varepsilon\in\Delta$ we let $v$
denote the vector whose components in the chosen basis are $v_M =
\bar\varepsilon\Gamma_M \varepsilon$.  Its Minkowskian norm $\eta^{MN}
v_M v_N$ is a quartic polynomial on spinors which is manifestly
$\Spin_{10,1}$-invariant.  It is possible to show that this norm is
negative semi-definite, whence $v$ is either time-like or null.
The orbit of a spinor in $\Delta$ for which $v$ is time-like
is $31$-dimensional and has isotropy $\SU_5$.  On the other hand, if
$v$ is null, the isotropy subgroup is a $30$-dimensional nonreductive 
Lie group which does {\em not\/} act trivially on the time direction.
It is a subgroup of the isotropy subgroup of $v$, which for a null
vector is isomorphic to $\Spin_9 \ltimes \RR^9$.  Indeed, as we will
see in the next section, the isotropy subgroup of such a spinor is
isomorphic to $\left(\Spin_7 \ltimes \RR^8\right)\times \RR$, where
$\Spin_7$ acts on $\RR^8$ according to the half-spin representation.
In terms of its action on $\MM^{10,1}$, $\Spin_7\subset\SO_8$ acts as
rotations in some $\EE^8 \subset \MM^{10,1}$, and $\RR^9$ acts via
null rotations.

\subsection{An exotic spinor isotropy group}

In this section we describe a certain 30-dimensional non-reductive
Lie subgroup $\eG \subset \Spin_{10,1}$ which leaves a spinor invariant.
We will exhibit its Lie algebra (and hence the Lie group itself)
inside the Clifford algebra $\Cl_{1,10}$ constructed above.

Consider a spinor $\varepsilon$ of the form
\begin{equation}\label{eq:spinor}
\varepsilon = \begin{pmatrix}
	      \psi_- \\ 0
	      \end{pmatrix}~,
\end{equation}
where $\psi_-$ is a negative chirality spinor of $\Cl_8$; that is,
$\Gamma'_9\psi_- = - \psi_-$.  It is easy to compute the isotropy
subalgebra $\fg \subset \so_{10,1}$ of $\varepsilon$ from the explicit
form of the generators of $\so_{10,1}$.  After a little bit of algebra
we obtain that the most general element of $\fg$ is given by
\begin{equation}\label{eq:isotropy}
\half a^{ij} \Sigma_{ij} + b^i \Sigma_{+i} + c \Sigma_{+\natural}~,
\end{equation}
where $\Gamma_+ = \Gamma_0 + \Gamma_9$, $b^i$ and $c$ are arbitrary
and $a^{ij}$ are such that $\half a^{ij} \Sigma'_{ij} \in \so_8$ is
in the isotropy subalgebra of $\psi_-$.  Because $\Spin_8$ acts
transitively on the unit sphere in both its spinor (as well as the
vector) representations, the isotropy subalgebras of every spinor are
conjugate, hence isomorphic.  This implies that the isotropy subgroup
of $\psi_-$ is a $\Spin_7$ subgroup: one which decomposes the
negative spinor representation of $\Spin_8$ but keeps the vector
representation irreducible.  This means that $\half a^{ij}\Sigma_{ij}$
belong to an $\so_7$ subalgebra of $\so_{10,1}$.  Computing the
Lie bracket of elements of the form \eqref{eq:isotropy}, and using
that $\Gamma_+$ squares to zero, we obtain
\begin{equation*}
\fg \cong \left ( \so_7 \ltimes \RR^8 \right) \times \RR~,
\end{equation*}
where $\RR^8$ is abelian and $\so_7$ acts on it as a spinor.  Notice
that $\RR$ is in the centre, and that $\RR^8$ is an abelian ideal,
whence this Lie algebra is not reductive.  Exponentiating inside
$\Cl_{1,10}$ we obtain a simply-connected 30-dimensional non-reductive
Lie subgroup $\eG \subset\Spin_{10,1}$ with the following structure
\begin{equation*}
\eG \cong \RR \times \left(\RR^8\rtimes \Spin_7\right)~.
\end{equation*}

\subsection{Invariants}

We now investigate the action of $\eG$ on $\MM^{10,1}$.  We will see
that it acts reducibly yet indecomposably.  It is convenient to
parametrise $\eG$, which topologically is homeomorphic to $\RR^9 \times
\Spin_7$, as follows:
\begin{equation*}
\Cl_{1,10} \supset \eG \ni g = \exp\left(c_\mu \Sigma_{+\mu}\right)\,
\sigma~,
\end{equation*}
where $\sigma\in\Spin_7$ and $\mu = (i,\natural)$, where $i$ runs
from $1$ to $8$.  Notice that the exponential only consists of two
terms because $\Gamma_+^2 = 0$:
\begin{equation*}
\exp\left( c_\mu \Sigma_{+\mu}\right) = 1 + c_\mu \Sigma_{+\mu} =
1 + \half c_\mu \Gamma_\mu\Gamma_+~.
\end{equation*}
The composition of group elements follows the standard semidirect
product structure:
\begin{equation*}
\exp\left( c_\mu \Sigma_{-\mu}\right)\,\sigma\,\exp\left( d_\mu
\Sigma_{-\mu}\right)\,\tau = \exp\left( \left(c_\mu + \sigma\cdot
d_\mu\right) \Sigma_{-\mu}\right)\,\sigma\tau~,
\end{equation*}
where $\sigma,\tau\in\Spin_7$ and $c_\mu, d_\mu\in\RR^9$.

\begin{table}[h!]
\centering
\setlength{\extrarowheight}{5pt}
\begin{tabular}{|c|>{$}l<{$}|}
\hline
Degree& \multicolumn{1}{c|}{Invariant forms}\\
\hline\hline
0 & 1\\
1 & e_+^*\\
2 & e_+^*\wedge e_\natural^*\\
5 & e_+^*\wedge\Omega\\
6 & e_+^*\wedge e_\natural^*\wedge\Omega\\
9 & e_+^*\wedge \vol_8 = e_+^*\wedge e_1^* \wedge \cdots \wedge
e_8^*\\
10& e_+^*\wedge e_\natural^* \wedge \vol_8\\
11& \vol_{10,1} = e_0^* \wedge e_1^* \wedge \cdots \wedge e_\natural^*
\\[5pt] \hline
\end{tabular}
\vspace{8pt}
\caption{$\eG$-invariant forms.  Here $e_M^*$ are canonical dual bases
to $e_M$, $\Omega$ is the Cayley form, and $\vol_8$ and $\vol_{10,1}$
are the eight- and eleven-dimensional volume forms, respectively.}
\label{tab:forms}
\end{table}

Let $g\in \eG$ be as above.  Its action on the basis $\{e_M\}$ is given
by
\begin{align}\label{eq:action}
g\cdot e_i &= \sigma\cdot e_i - \sigma^{-1} \cdot c_i\,e_+\notag\\
g\cdot e_9 &= e_9 + c_\mu e_\mu - \half |c|^2 e_+\notag\\
g\cdot e_\natural &= e_\natural - c_\natural e_+\notag\\
g\cdot e_0 &= e_0 - c_\mu e_\mu + \half |c|^2 e_+\notag\\
\therefore\qquad g\cdot e_+ &= e_+\notag\\
\therefore\qquad g\cdot e_- &= e_- - 2 c_\mu e_\mu + |c|^2 e_+~.
\end{align}
Notice that for nonzero $c_\mu$ exactly one null direction is left
invariant, so that the transformation is a null rotation.  From these
formulae above one can determine the space of $\eG$-invariant forms.
The results are summarised in Table \ref{tab:forms}.

Decomposing the spinor representation $\Delta$ under $\Spin_7$, there
are precisely two linearly independent $\Spin_7$-invariant spinors.
The null rotations in $\RR^9$ preserve only one of them.  Therefore
$\eG$ leaves invariant exactly one spinor (up to scale)---the spinor
$\varepsilon$ in \eqref{eq:spinor}, where $\psi_-$ is the
$\Spin_7$-invariant spinor in that representation.  This means that
any configuration of $m$ $\M$-branes whose tangent planes are
$\eG$-related will be supersymmetric, provided that $\varepsilon$
belongs to $\Delta(\pi)$ in the first place.


We conclude this section with a useful fact about null rotations.

\begin{lem}
If a spinor in $\Delta$ is annihilated by a (nontrivial) null
rotation, then it is annihilated by all null rotations.
\end{lem}

\begin{proof}
Let $\varepsilon = (\psi~\chi)^t$ be a spinor in $\Delta$, and let $N
= c_\mu \Sigma_{+\mu}$ be an infinitesimal null rotation.  In terms of
the above realisation, we have
\begin{equation*}
N = \begin{pmatrix}
    - c_i\, \Gamma'_i\, \Pi_+ & c_\natural\, \Pi_-\\[3pt]
    - c_\natural\, \Pi_+ & c_i\, \Gamma'_i\, \Pi_-
    \end{pmatrix}~,
\end{equation*}
where $\Pi_\pm = \half ( \1 \pm \Gamma'_9)$ are the chiral projectors
for $\Cl_8$.  Therefore $N \cdot \varepsilon = 0$ if and only if
\begin{equation}\label{eq:nullrotn}
c_\natural \chi_- = c_i \Gamma'_i \psi_+\qquad \text{and}\qquad
c_\natural \psi_+ = c_i \Gamma'_i \chi_-~.
\end{equation}
Iterating these equations, we see that
\begin{equation*}
|c|^2\psi_+ = 0 \qquad\text{and}\qquad |c|^2 \chi_- = 0~.
\end{equation*}
Therefore either $c_\mu =0$ or $\psi_+ = \chi_- = 0$, in which case
\eqref{eq:nullrotn} is satisfied for all $c_\mu$.  In other words,
the kernel of any nontrivial infinitesimal null rotation $c_\mu
\Sigma_{+\mu}$ coincides with the kernel of $\Gamma_+$.
\end{proof}

\subsection{More on vectors and spinors}

In this section we collect one final result we shall need in order to
treat the multiple intersection problem.  We state the result in a
little bit more generality than is needed.

\begin{prop}
Let $\varepsilon\in\Delta(\pm\pi)$ for a fixed plane $\pi$.  Then the
vector $v$ associated to $\varepsilon$ lies in $\pi$.
\end{prop}

\begin{proof}
Let $\pi$ be a nondegenerate plane and $\pi^\perp$ its
orthocomplement.  Any vector $v$ splits uniquely as $v_\top + v_\bot$,
where $v_\top \in \pi$ and $v_\bot \in\pi^\perp$.  It is easy to show
that $v_\top \cdot \pi = - \pi \cdot v_\top$ and that $v_\bot \cdot
\pi = \pi \cdot v_\bot$.  Let us introduce projectors $\PP_\pm = \half
(\1 \pm \pi)$.  Let $\varepsilon$ belong to $\Delta(\pm\pi)$, and let
us compute the components $\bar\varepsilon \cdot v \cdot \varepsilon$
of the vector $v$.  From the properties of the charge conjugation $C$
in $\Cl_{1,10}$,
\begin{equation*}
v^t \cdot C = - C \cdot v~,
\end{equation*}
we deduce that
\begin{equation*}
\pi \cdot C = - C \cdot \pi \implies \PP_\pm \cdot C = C \cdot
\PP_\mp~.
\end{equation*}
Therefore, since $\PP_\pm \cdot \varepsilon = \varepsilon$,
\begin{align*}
\bar\varepsilon \cdot v \cdot \varepsilon &=
\overline{\PP_\pm\cdot\varepsilon} \cdot v \cdot \varepsilon\\
&= \bar\varepsilon \cdot \PP_\mp \cdot \left( v_\top +
v_\bot\right) \cdot \varepsilon\\
&= \bar\varepsilon \cdot \left( v_\top \cdot \PP_\pm + v_\bot
\cdot \PP_\mp \right) \cdot \varepsilon\\
& = \bar\varepsilon \cdot v_\top \cdot \varepsilon~,
\end{align*}
so that $v = v_\top$.
\end{proof}

\section{Multiple intersections}

In this section we discuss multiple intersections.  The idea of the
construction of supersymmetric intersections of branes is very simple.
Suppose that two coincident branes have tangent plane $\pi$.  This
configuration preserves one half of the supersymmetry, corresponding
to the asymptotic values $\varepsilon$ which belong to
$\Delta(\pi)$---i.e., which satisfy equation \eqref{eq:fundsusy}.  Now
let $L$ denote any Lorentz transformation in $\SO^0_{10,1}$, and let
$\widehat{L}$ denote a lift to $\Spin_{10,1}$. If the lift can be
chosen to lie in the isotropy subgroup of the spinor, so that
\begin{equation*}
\widehat{L}\cdot \varepsilon = \varepsilon~,
\end{equation*}
then the Lorentz-transformed brane $L\pi$ also satisfies
\begin{equation*}
L\pi \cdot \varepsilon = \varepsilon~.
\end{equation*}
This means that the brane configuration with tangent planes $\pi$ and
$L\pi$ is supersymmetric.  We must therefore consider subgroups $G$ of
$\Spin_{10,1}$ which leave invariant a number of spinors in
$\Delta(\pi)$.   Then these spinors will also belong to $\Delta(g\pi)$
for all $g\in G$.  The problem is therefore to classify all such
subgroups $G$ and compute the fraction of the supersymmetry which is
preserved by a generic configuration consisting of $G$-related branes.

\subsection{$G$-relatedness}

To make this precise, let us adapt the definition of $G$-relatedness
given in Part~I to our more general situation.  Let
$G((5,1),\MM^{10,1})$ denote the grassmannian of oriented
time-oriented $(5,1)$-planes in $\MM^{10,1}$.  It is acted on
transitively by $\SO^0_{10,1}$ with isotropy
$\SO^0_{5,1}\times\SO_5$. A given subgroup $G\subset\Spin_{10,1}$ acts
on $G((5,1),\MM^{10,1})$ by restricting the action of $\SO^0_{10,1}$
to the subgroup to which $G$ gets mapped under the canonical covering
map $\Spin_{10,1}\to\SO^0_{10,1}$.  We can therefore consider the
decomposition of the grassmannian into $G$-orbits.

\begin{dfn}
Let $G\subset\Spin_{10,1}$ and let $\{\pi_i\}$ be $m$ oriented
time-oriented $(5,1)$-planes in $\MM^{10,1}$.  We say that they are
{\em $G$-related\/}, if they all lie in the same $G$-orbit and
furthermore $G$ is the {\em smallest\/} such subgroup of
$\Spin_{10,1}$.
\end{dfn}

In Part~I we analysed in detail the case $G\subset\Spin_{10}$.  In
\cite{AFS-cali} we proved that a configuration of $m$ static
intersecting branes is supersymmetric if and only if the tangent
planes are $G$-related, where $G\subset\SU_5$ or $G\subset\Spin_7$ or
both, whence $G\subset\SU_4$.  This $\Spin_7$ subgroup is in fact the
intersection with $\Spin_{10}$ of the exotic isotropy subgroup
$(\Spin_7\ltimes\RR^9) \subset\Spin_{10,1}$.  Indeed, as we will see
presently, a configuration of $m$ intersecting branes in motion is
supersymmetric if and only if their tangent planes are $G$-related,
where $G = K \ltimes N$ and where $K \subset\Spin_7$ and $N\subset
\RR^9$.  In fact, we will see that every supersymmetric intersection
of $\M5$-branes in motion will correspond to a configuration of
Cayley planes in an eight-dimensional euclidean subspace of
$\MM^{10,1}$, which have been null-rotated in the remaining three
dimensions.

In Part~I we also proved a lower bound for the fraction $\nu$ of the
supersymmetry preserved by a configuration of $G$-related branes in
terms of the action of the group $G\subset\Spin_{10}$ in the spinor
representation $\Delta$.  We also conjectured that this lower bound
was in fact saturated.  In the present case, where $G \subset \eG$,
the situation is more complicated.  Clearly the fraction $\nu$ of the
supersymmetry which is preserved by a configuration of $G$-related
planes will be larger than or equal to the fraction $\nu_G$ defined by
\begin{equation*}
32 \nu_G = \dim\left( \Delta(\pi) \cap \Delta^G\right)~,
\end{equation*}
where $\Delta^G$ is the space of $G$-invariant spinors.
For $G\subset\Spin_{10}$, we proved in Part~I that $\dim \left(
\Delta(\pi) \cap \Delta^G\right) = \half \dim\Delta^G$; but for $G$
containing null rotations, this is not always the case.  Let us
analyse this now, and in so doing reduce the problem to one concerning
$4$-planes in eight dimensions.

\subsection{An equivalent eight-dimensional problem}
\label{sec:8dim}

It is not hard to show that the supersymmetric configurations of
intersecting branes in motion consist of null rotated Cayley planes.
In fact, suppose that we consider $G$-related planes where $G=
K\ltimes N \subset\Spin_7 \ltimes \RR^9$, where $N$ is not the trivial
group. Let $\pi$ be one of the planes, and let $v$ denote an
$N$-invariant null vector.  By the Proposition, $v$ belongs to $\pi$.
Because $\pi$ is nondegenerate, $\pi$ also contains a complementary
null vector $v'$, whence we can write $\pi = v \wedge v'
\wedge \zeta$ where $\zeta$ is a $4$-plane in $\spn\{v,v'\}^\perp
\cong \EE^9$.  Now, $\Spin_7$ leaves a direction $u$ in this $\EE^9$
fixed.  Therefore $K\subset\Spin_7$ moves $\zeta$ only in an
eight-dimensional subspace $\spn\{v,v',u\}^\perp\cong \EE^8$.

We can see this a little bit more explicitly.  We want to study
$\Delta^G \cap \Delta(\pi)$ and compute its dimension in terms of
$G$, in the case where $G = K \ltimes N \subset \Spin_7 \ltimes
\RR^9$.  We take $\pi = e_+ \wedge e_- \wedge \zeta$, with $\zeta =
e_1\wedge e_3 \wedge e_5 \wedge e_7$.  By the Proposition, the null
vector associated to this null rotation belongs to $\pi$.  Using the
Lorentz invariance on $\pi$, we can assume that it is $e_+$.  Since
$G$ contains a nontrivial null rotation, we can first consider the
subspace $\Delta'\subset\Delta$ defined by
\begin{equation*}
\Delta' = \{\varepsilon \in \Delta \mid \Gamma_+ \cdot \varepsilon =
0\}~.
\end{equation*}

By (the proof of) the Lemma, $\Delta' = \Delta^N$.  It is a
$16$-dimensional real subspace of $\Delta$.  $K$ preserves $\Delta'$
because it is contained in $\Spin_8$, which commutes with $\Gamma_+$.
Under $\Spin_8$, $\Delta'$ breaks up as two irreducible
representations: one of each chirality. In fact, from the proof of the
Lemma, it follows that in the chosen realisation, a spinor
$\varepsilon$ belongs to $\Delta'$ if and only if it has the form
\begin{equation*}
\varepsilon = \begin{pmatrix}
	      \psi_- \\ \chi_+
	      \end{pmatrix}~.
\end{equation*}
From the explicit expression for $\pi$, $\varepsilon \in\Delta(\pi)$
if and only if
\begin{equation*}
\zeta \cdot \psi_- = \psi_-\qquad\text{and}\qquad \zeta \cdot \chi_+ =
\chi_+~.
\end{equation*}
In other words, if we make a spinor $\varphi = (\psi_-~\chi_+)^t$ of
$\Cl_8$, then $\varepsilon\in\Delta(\pi)$ if and only if $\varphi
\in\Delta'(\zeta)$ in the obvious notation.  In other words,
the subspaces $\Delta^G \cap \Delta(\pi)$ and $(\Delta')^K \cap
\Delta'(\zeta)$ of $\Delta$ agree.

It is therefore possible to work with $\Delta'$ abstractly as the
irreducible representation of $\Cl_8$ and to consider $4$-planes
$\zeta$ in $\EE^8$.  As shown in \cite{LM} (see also
\cite{Harvey,DadokHarvey-spinors,AFS-cali}), a $4$-plane $\zeta$ in
$\EE^8$ which satisfies
\begin{equation*}
\zeta \cdot \varphi = \varphi~,
\end{equation*}
is a Cayley plane; that is, it is calibrated by the Cayley form built
out of $\varphi$ by squaring.  In other words, the $G$-related planes are
$K$-related Cayley planes which are then null-rotated by $N$.

The most generic situation results from $K=\Spin_7$ and $N=\RR^9$.
Then we simply have null-rotated Cayley planes.  This is a
$17$-di\-men\-sion\-al orbit inside the $30$-dimensional grassmannian
of $(5,1)$-planes.  Such configurations will generically preserve
$\frac{1}{32}$ of the supersymmetry, because there is at most one
spinor in $\Delta'$ which is left invariant by a $\Spin_7$ subgroup.
Configurations with a larger fraction are possible provided that we
restrict to planes which live in progressively smaller subspaces of
the grassmannian.  If $\cup_{i=1}^m \zeta_i$ are $K$-related, then the
resulting brane configuration preserves a fraction $\nu$ of the
supersymmetry which obeys $\nu \geq \nu_K$, where $\nu_K$ is given by
\begin{equation*}
\nu_K = \tfrac{1}{32} \dim\left( \Delta'(\zeta) \cap
(\Delta')^K\right)~.
\end{equation*}
In Part~I we showed that the group-theoretical fraction was directly
related to the dimension of group invariant spinors.  In the case of
non-static branes treated here, the dependence of $\nu_K$ on $K$ is
more subtle, as we now explain.

\subsection{The group-theoretical fraction $\nu_K$}
\label{sec:fraction}

We would like to compute the dimension of the subspace $\Delta(\pi)
\cap \Delta^G \subset \Delta$, where $G = K\ltimes N$.  In the
previous section we saw how this computation can be rephrased in terms
of $K$-related Cayley planes in eight dimensions.  This was achieved
by first taking care of the null rotations and thus reducing the
problem to one in lower dimension.  Here we will invert the order and
first take care of the $K$-invariance.  Therefore let $\Delta^K
\subset \Delta$ denote the space of $K$-invariant spinors in
$\Delta$.  $K\subset\Spin_8$ and, as we saw in
\eqref{eq:spin8branching}, $\Delta$ breaks up under
$\Spin_{10,1}\supset\Spin_8$ as two copies of each of the half-spin
representations.  Therefore $\Delta^K$ is even-dimensional.  As shown
in (the proof of) the Lemma, $N$-invariant spinors are those spinors
which are annihilated by $\Gamma_+$.  Because $\Gamma_+$ is
$K$-invariant, $\Gamma_+$ maps $\Delta^K$ to itself. By the same
token, so does $\Gamma_-$.  Because
\begin{equation*}
\Gamma_+ \Gamma_- + \Gamma_- \Gamma_+ = - 4 \1~,
\end{equation*}
reasons similar to those from which we deduced \eqref{eq:Hodgedec}
also yield the following decomposition of $\Delta^K$:
\begin{equation*}
\Delta^K = \Delta^K_+ \oplus \Delta^K_-~,
\end{equation*}
where $\Delta^K_\pm = \Delta^K \cap \ker\Gamma_\pm = \Gamma_\pm
\Delta^K$.  Similarly, it follows that $\Delta^K_\pm$ are isomorphic
subspaces, the isomorphisms being given by $\Gamma_\pm$.  In
particular, $\dim\Delta^K_\pm = \half \dim\Delta^K$.  Because
$\Delta^G = \Delta^K_+$, we see that
\begin{equation*}
\dim\Delta^G = \half \dim\Delta^K~.
\end{equation*}

Now we intersect $\Delta^G$ with $\Delta(\pi)$.  Let
$\left(\Delta^K\right)^\pm = \Delta^K \cap \Delta(\pm\pi)$.  Similarly
we can define $\left(\Delta^K_\pm\right)^\pm$ in the obvious way,
where the signs are not correlated.  This gives rise to a fourfold
decomposition of $\Delta^K$:
\begin{equation*}
\Delta^K = \left(\Delta^K_+\right)^+ \oplus \left(\Delta^K_+\right)^-
\oplus \left(\Delta^K_-\right)^+ \oplus \left(\Delta^K_-\right)^-~.
\end{equation*}
By the Proposition, $e_\pm$ belong to $\pi$, whence $\Gamma_\pm \cdot
\pi = - \pi \cdot \Gamma_\pm$.  This means that $\Gamma_\pm$ restrict
to isomorphisms
\begin{equation*}
\Gamma_\pm : \left(\Delta^K_\mp\right)^+ \to
\left(\Delta^K_\pm\right)^-\qquad\text{and}\qquad
\Gamma_\pm : \left(\Delta^K_\mp\right)^- \to
\left(\Delta^K_\pm\right)^+~.
\end{equation*}
Letting $d_\pm^\pm = \dim \left(\Delta^K_\pm\right)^\pm$, we have that
$d_+^+ = d_-^-$ and that $d_+^- = d_-^+$, whence
\begin{equation*}
\dim\Delta^K = 2\,d_+^+ + 2\,d_+^-~.
\end{equation*}
In other words, since $\left(\Delta(\pi)\cap\Delta^G\right) =
\left(\Delta^K_+\right)^+$, its dimension is given by
\begin{equation*}
d_+^+  = \half \dim\Delta^K - d_+^- \leq \half \dim\Delta^K~.
\end{equation*}
It will be convenient in what follows to introduce a rational number
$\varrho_K$ defined implicitly by
\begin{equation*}\label{eq:ratio}
d_+^+ = \varrho_K \dim\Delta^K~.
\end{equation*}

In contrast with the case of static branes, where $G\subset
\Spin_{10}$, we cannot compute $d_+^+$ in a uniform manner, for it
seems to depend on other properties of the group besides its
invariants on the spinor representation; that is, the ratio
$\varrho_K$ depends nontrivially on $K$.  For example, we saw already
that for $m{=}2$ branes, $d_+^+ = d_-^+$, whence $\varrho_K =
\tfrac{1}{4}$.  On the other hand, in some explicit examples we have
computed (and which will be discussed briefly below) we also find
cases in which $d_+^- = 0$ so that $\varrho_K = \half$, and even some
cases for which $\varrho_K$ takes less obvious values $\tfrac18$,
$\tfrac16$, $\tfrac15$, $\tfrac{3}{10}$, $\tfrac13$, and $\tfrac38$.
This seems to indicate the need for a case-by-case analysis.  We now
discuss some explicit examples.

\subsection{Some examples and their geometries}

In this section we list some examples of $G$-related branes in motion
which we have worked out explicitly.  These examples are summarised in
Table \ref{tab:geometries}.  Many of the necessary calculations have
been performed infinitesimally (i.e., using their Lie algebras) using
{\em Mathematica\/}.\footnote{Details of the calculations can be
obtained by email from the authors.  They will be made public via our
web pages at a later date.}

\begin{table}[h!]
\centering
\setlength{\extrarowheight}{5pt}
\begin{tabular}{|*{6}{>{$}c<{$}|}}
\hline
\multicolumn{1}{|c|}{Codim.}&
\multicolumn{1}{c|}{Group}&
\multicolumn{1}{c|}{Fraction}&
\multicolumn{1}{c|}{Ratio}&
\multicolumn{1}{c|}{Isotropy}&
\multicolumn{1}{c|}{Geometry}\\
d&
K&
\nu_K&
\varrho_K&
H&
K/H\\[3pt]
\hline\hline
&\Spin_7  &\tfrac{1}{32} & \half &(\SU_2)^3/\ZZ_2 & \text{Cayley} \\
& \SU_4 & \tfrac{1}{32} & \tfrac14 & \SO_4 & \slg_4\\
& \Sp_2 & \tfrac{1}{32} & \tfrac16 &\U_2 & \clg_2 \\
& \Sp_1 \times \Sp_1 & \tfrac{1}{32} & \tfrac18 & \Sp_1 &
(3,1)~\text{in \cite{DadokHarveyMorgan}}\\
& \SU_4 & \tfrac{1}{16} & \half &\mathrm{S}(\U_2\times\U_2) & \CC_2\\
4& \Sp_2 & \tfrac{1}{16} & \tfrac13 &\U_2 & \clg_2 \\
& \Sp_1 \times \Sp_1 & \tfrac{1}{16} & \tfrac14 & \U_1\times\U_1 &
\CC_1\times\CC_1\\
&\Sp_1 & \tfrac{1}{16} & \tfrac15 & \U_1 & (3,2)~\text{in
\cite{DadokHarveyMorgan}}\\
& \Sp_2 & \tfrac{3}{32} & \tfrac12 &\Sp_1\times\Sp_1 & \HH_1 \\
& \Sp_1 \times \Sp_1 & \tfrac{3}{32} & \tfrac38 & \Sp_1 &
(3,1)~\text{in \cite{DadokHarveyMorgan}}\\
& \Sp_1 & \tfrac{3}{32} & \tfrac{3}{10} &\U_1 & (3,2)~\text{in
\cite{DadokHarveyMorgan}}\\
& \U_1 & \tfrac{3}{32} & \tfrac14 &\{1\} & (3,3)~\text{in
\cite{DadokHarveyMorgan}}\\[5pt]
\hline
3& G_2 &\tfrac{1}{16} & \half &\SO_4 & \text{Associative}\\
& \SU_3 &\tfrac{1}{16} & \tfrac14 & \SO_3 & \slg_3\\[5pt]
\hline
2&\SU_3 &\tfrac{1}{8} & \half & \mathrm{S}(\U_2\times\U_1) & \CC_1\\
& \SU_2 &\tfrac{1}{8} & \tfrac14 & \SO_2 &
\CC_1~\text{or}~\slg_2\\[5pt]
\hline
\end{tabular}
\vspace{8pt}
\caption{Some of the geometries associated with supersymmetric
configurations of multiply intersecting branes in motion.  The table
has been compiled in terms of the equivalent eight-dimensional problem
discussed in Section~\ref{sec:8dim}.}
\label{tab:geometries}
\end{table}

To simplify the table we have used the equivalent eight-dimensional
description of the branes in motion as Cayley planes which have been
null-rotated.  Therefore for branes in motion which are $G$-related,
where $G=K\ltimes N$, we have listed the group $K$ together with its
codimension, fraction and the ratio $\varrho_K$ defined in equation
\eqref{eq:ratio}.  We have also listed the isotropy subgroup $H\subset
K$ of the reference $4$-plane $\zeta$, and the geometry of the
resulting grassmannian $K/H$.  This is a sub-grassmannian of the
Cayley grassmannian, whence it does not take into account the null
rotations.  The true grassmannian of $G$-related planes is now a
homogeneous bundle over $K/H$ with fibre $\RR^5$, where the $\RR^5$
factor corresponds to those null rotations defined by vectors in the
orthogonal plane $\pi^\perp$.  Notice that the dependence of $\nu_K$
on the subgroup $K$ is subtle, since we find different values of
$\varrho_K$ for different yet isomorphic subgroups $K$.

Notice also that all the geometries which appear have already been
discussed in Part~I and hence will not be discussed further here,
except for two brief remarks.  First, we would like to bring to the
attention of the reader the fact that in this table, and in contrast
with the similar table in Part~I, isomorphic subgroups $K$ giving rise
to isomorphic geometries now yield configurations with different
fractions.  This does not mean that the dimension of the subspace of
$K$-invariant spinors is different, for this only depends on the
isomorphism class of $K$.  Instead, as mentioned above, it is the
intersection of this subspace with $\Delta(\pi)$ which depends subtly
on $K$.  Finally, let us remark that most of the entries in the table
distinguished by having $\varrho_K\neq\half$, correspond to different
subgroups than the ones giving rise to the same geometries in the
similar table in Part~I.

\section{Summary and Outlook}

In this paper we have obtained new supersymmetric configurations of
intersecting branes, consisting of branes which are in relative motion
to each other.  This work completes the classification of
supersymmetric configurations of two intersecting $\M5$-branes,
started by Ohta \& Townsend in \cite{OhtaTownsend}.  As shown in
Part~I for the static configurations and in the present paper for the
rest, each configuration is characterised by (the maximal torus of) a
subgroup $G$ of $\Spin_{10,1}$ contained in the isotropy of a spinor.
The fraction of the supersymmetry which is preserved is given by
$\tfrac{1}{32}$ times the dimension of the subspace of $\Delta$
consisting of those spinors invariant under $\TT(G)$, or when $G = K
\ltimes N$, under $\TT(K) \times \RR$.  These results are summarised
in Table \ref{tab:2M5branes}.

\begin{table}[h!]
\centering
\setlength{\extrarowheight}{5pt}
\begin{tabular}{|>{$}c<{$}|>{$}l<{$}|>{$}l<{$}|}
\hline
\text{Fraction}& \multicolumn{2}{c|}{Rotation subgroups for}\\
\nu & \multicolumn{1}{c|}{static branes} & \multicolumn{1}{c|}{branes
in motion}\\
\hline\hline
\tfrac{1}{32} & \SU_5 & \Spin_7 \circeq \SU_4\\
\tfrac{1}{16} & \SU_2\times\SU_3\quad\Spin_7\circeq\SU_4 &
\Sp_2\circeq(\Sp_1)^2\quad G_2 \circeq\SU_3\\
\tfrac{3}{32} & U_1\times\SU_2 & \Sp_1\circeq U_1\\
\tfrac{1}{8} & U_1\quad\Sp_2\circeq(\Sp_1)^2\quad
G_2\circeq\SU_3 & \ZZ_2\quad\SU_2\\
\tfrac{5}{32} & \ZZ_6 & \\
\tfrac{3}{16} & \Sp_1 \circeq\U_1 & \\
\tfrac{1}{4} & \ZZ_2\quad\SU_2 & \{1\}\\
\tfrac{1}{2} & \{1\} & \\[5pt] \hline
\end{tabular}
\vspace{8pt}
\caption{All possible supersymmetric configurations of two
$\M5$-branes, and the corresponding subgroups of $\Spin_{10,1}$.  In
the case of branes in motion, only the rotation subgroup has been
indicated---the null rotations remaining implicit.  The relation
$\circeq$ denotes subgroups having the same maximal torus.  Some
finite groups discussed briefly in Part~I have been included for
completeness.}
\label{tab:2M5branes}
\end{table}

One can construct each of these configurations starting with two
coincident fivebranes and rotate by an element of the maximal torus of
a subgroup $G$ of $\Spin_{10}$.  If $G$ is actually contained in
$\Spin_8$ then one can also perform a null-rotation in the
($2+1$)-dimensional subspace left invariant by $\Spin_8$.  This gives
rise to configurations in which the branes are moving relative to each
other.

The classification of multiple intersecting brane configurations is
more complicated and a complete solution is still lacking.  In
\cite{AFS-cali,AFS-groups} as well as in the present paper, we have
shown that a configuration preserves some supersymmetry provided that
the branes are $G$-related where $G$ is contained in the isotropy of a
spinor which obeys equation \eqref{eq:fundsusy} relative to one of the
branes.  The possible groups $G$ fall into two classes, depending on
whether $G$ is contained or not in $\Spin_{10}$.  If $G \subset
\Spin_{10}$ then $G$-related branes are static relative to each other
since they all share the same time-like direction.  If $G$ is not
contained in $\Spin_{10}$ we have shown that $G$ contains null
rotations, so that $G = K \ltimes N$, where $K \subset\Spin_8$ and $N$
contains null rotations.  Configurations of $G$-related fivebranes can
be understood as $K$-related euclidean fourbranes in eight dimensions
which have then been null rotated.  These configurations share the
same null direction.   In Part~I we restricted ourselves (following
\cite{OhtaTownsend}) to static brane configurations and derived a
(conjecturally exact) lower bound $\nu_G$ for the fraction $\nu$ of
the supersymmetry preserved by a configurationof $G$-related branes
with $G\subset\Spin_{10}$, in terms of the dimension of the space of
$G$-invariant spinors.  For $G = K \ltimes N$ containing null
rotations, we have exhibited a lower bound $\nu_K$ for $\nu$, but as
shown by explicit examples, the precise relationship between $\nu_K$
and $K$ is more subtle and a simple uniform expression for all $K$
would be desirable.  The intricate dependence of the fraction on the
subgroup $K$ does not discard the possibility of finding fractions
$\nu$ in this way which have not been encountered before.  For the
case of branes in motion, where $\nu\leq \tfrac14$, the only fractions
which have not appeared so far are $\tfrac{5}{32}$, $\tfrac{3}{16}$
and $\tfrac{7}{32}$; although all but the latter have occured for
static branes.

In this series of papers we have so far focused mostly on
$\M5$-branes; yet similar results are also valid for other types of
branes both in $\M$-theory as in ten-dimensional superstring theories.
In fact, most if not all supersymmetric brane configurations in
superstring theory can be related by dualities to supersymmetric brane
configurations in $\M$-theory, and hence the classification of
$\M$-theory brane configurations would in principle also classify
those.  At the same time, it is not enough to classify configurations
featuring only one type of $\M$-brane: in order to take into account
all BPS states it is necessary also to consider configurations
consisting of branes of different types, as in the following example.

Let $\varepsilon$ be a spinor of the form
\begin{equation*}
\varepsilon = \begin{pmatrix}
	     \psi_- \\ \chi_+
	     \end{pmatrix}~.
\end{equation*}
From the explicit expression for the $\Gamma$ matrices given in
\eqref{eq:gammas}, we can see that
\begin{equation}\label{eq:Mwave}
\left( e_0 \wedge e_9\right) \cdot \varepsilon = \varepsilon~,
\end{equation}
which is the algebraic statement corresponding to the fact that the
$\M$-wave solution preserves $\half$ of the supersymmetry
\cite{Townsend-Malgebra}.  Similarly, we have that for spinors of the
form
\begin{equation*}
\varepsilon = \begin{pmatrix}
	     \psi_- \\ \chi_-
	     \end{pmatrix}
\end{equation*}
the following holds:
\begin{equation}\label{eq:M2brane}
\left( e_0 \wedge e_9 \wedge e_\natural \right) \cdot \varepsilon =
\varepsilon~,
\end{equation}
which now says that an $\M2$-brane stretched along the $09\natural$
directions is supersymmetric.  Now let $G = K \ltimes N \subset\eG$
and perform a $G$-transformation to each of the above supersymmetric
brane solutions: wave \eqref{eq:Mwave} and membrane
\eqref{eq:M2brane}.  Since $G$ stabilises a spinor $\varepsilon$ of
the form \eqref{eq:spinor}, the new configurations consisting of the
original and the transformed branes is supersymmetric but each
preserving $\tfrac{1}{4}$ of the supersymmetry.  Notice then that for
spinors of the form \eqref{eq:spinor} we can have configurations
containing both $\M$-waves and $\M2$-branes in motion which preserve
$\tfrac14$ of the supersymmetry.  Similarly it is not difficult to
construct other configurations involving also $\M5$-branes.  We hope
to return to a detailed discussion of the general problem in a future
publication.

\section*{Acknowledgements}

It is a pleasure to thank Robert Bryant for his helpful correspondence
and for sending us his unpublished notes \cite{Bryant-spinors}, and
Yolanda Lozano for useful discussions at the early stages of this
work.  Part of this work was done while JMF was visiting the
Department of Mathematics of Boston University, and he would like to
thank Takashi Kimura for arranging the visit and for hospitality.
Finally, BSA and SS are supported by PPARC Postdoctoral Fellowships,
JMF by an EPSRC PDRA, and BS by an EPSRC Advanced Fellowship, and we
would like to extend our thanks to the relevant research councils for
their support.

%
%

\end{document}